\documentclass[11pt]{article}   %\usepackage{dvisrc} %[30.10.2000]
\makeatletter
\def\draft{\textheight=10.5truein \textwidth=7.5truein \parindent=8pt
           \voffset=-1truein \topmargin=0Truein
           \ifcase \@ptsize \hoffset=-1.5truein \or \hoffset=-1.35truein
                        \or \hoffset=-1.15truein \fi}
\def\quality{\textheight=240mm \textwidth=160mm \topmargin=0Truein
             \ifcase \@ptsize \hoffset=-23mm
                     \or \hoffset=-20mm \or \hoffset=-15mm \fi}
\makeatother
\quality

\def\beq#1#2{\begin{equation} \label{#1} #2 \end{equation}}
\def\function#1{\left\{\!\!\!\begin{array}{ll} #1 \end{array} \right.}

\def\CR{$$ $$}    \def\n{\noindent}
  \def\IL{{\bf L}}  
\def\IN{\hbox{I\kern-.2em\hbox{N}}}
\def\IR{\hbox{\rm I\kern-.2em\hbox{\rm R}}}
\def\IZ{\hbox{{\rm Z}\kern-.3em{\rm Z}}}
\def\ep{\varepsilon}  \def\phi{\varphi}   \def\la{\lambda}

\def\bX{{\bf X}} \def\c#1{{{#1}^*}}
\def\map{T} \def\v{{\rm V}}  \def\rel{\vdash}  \def\Ee{e}
\def\Reg{{\rm Reg}} \def\Per{{\rm Per}}  \def\Free{{\rm Free}}
\def\om{o}  \def\bC{{\bf C}}

\def\proof{\smallskip \noindent {\bf Proof. \ }}
\newcommand\filledsquare{\ \vrule width 1.5ex height 1.2ex}  %filled square
\def\qed{\hfill\filledsquare\linebreak\smallskip\par}

\def\thname{Theorem}     \def\lmname{Lemma}      \def\prname{Proposition}
\def\dfname{Definition}  \def\crname{Corollary}  \def\rmname{Remark}

\newtheorem{theorem}{\thname}[section]   %Numbering: Theorem--Other section
\newtheorem{lemma}{\lmname}[section]     %{lemma}[theorem]{Lemma}   section
\newtheorem{proposition}[lemma]{\prname} %lemma
\newtheorem{corollary}[lemma]{\crname}   %lemma

\newtheorem{dftn}{\dfname}[section]
\newtheorem{rmrk}[lemma]{\rmname}
             \catcode`@=11 \@addtoreset{equation}{section} \catcode`@=12
\begin{document}
%%%--------------------------------------------------------%%%%
\title{Dynamics of traffic jams: order and chaos}
\author{Michael Blank\thanks{This research has been partially
                             supported by CRDF RM1-2085,
                             INTAS 97-11134 and RFBR grants.}
       \\ \\
        Russian Ac. of Sci., Inst. for
        Information Transmission Problems, \\
        B.Karetnij 19, 101447, Moscow, Russia, blank@iitp.ru}
\date{January 28, 2001}%{\today} [final version for MMJ]}
\maketitle

\begin{abstract} By means of a novel variational approach we
study ergodic properties of a model of a multi lane traffic flow,
considered as a (deterministic) wandering of interacting
particles on an infinite lattice. For a class of initial
configurations of particles (roughly speaking satisfying the Law
of Large Numbers) the complete description of their limit (in
time) behavior is obtained, as well as estimates of the transient
period. In this period the main object of interest is the
dynamics of `traffic jams', which is rigorously defined and
studied. It is shown that the dynamical system under
consideration is chaotic in a sense that its topological entropy
(calculated explicitly) is positive. Statistical quantities
describing limit configurations are obtained as well.
\end{abstract}

\bigskip%
\n{\bf Keywords}: traffic flow, dynamical system, variational
principle, topological entropy.

\bigskip%
\n{\bf AMS Subject Classification}: Primary 37B99; Secondary
37B15, 37B40, 37A60, 60K.

%\bigskip
\section{Introduction}\label{section-intro}

Despite a self evident practical importance of the analysis of
traffic flows and a relatively long history of attempts of their
scientific treatment (going back to the fifties) only recently
(in the end of nineties) reasonable mathematical models of
traffic flows and method to study them were obtained. Previous
attempts were based on ideas borrowed from such classical fields
of physics as mechanics and hydrodynamics. Not going into details
of a qualitative and quantitative comparison of the hydrodynamic
type models with practice (which one can find, for example, in a
recent review \cite{Sch} and references therein), we consider the
following practical observation. It turns out that going by foot
in a slowly moving crowd it is faster to go against the ``flow''
than in the same direction as other people go. A mathematical
model describing this effect in the case of the one lane traffic
was introduced in \cite{Bl-car}. A standard probabilistic model
of a diffusion of particles against/along the flow clearly
contradicts to this observation, which very likely indicates a
very special (nonrandom) intrinsic structure of the flow in this
case. The main aim of the present paper is to study how this
structure emerges from arbitrary (random) initial configurations
in a simple model of the multi lane traffic flow.

A recent progress in the analysis of traffic flows was due to the
introduction of discrete (in time and in space) cellular automata
models of the one lane traffic flow in \cite{NS,NH} and later
studied by many authors (see \cite{BF} for review and further
references). Various approaches starting from the mean-field
approximation \cite{KS} to combinatorial techniques and
statistical mechanics methods \cite{F} were used in their
analysis. All these models were based on the idea to describe the
dynamics in terms of deterministic or random cellular automata
(see results about stochastic models in
\cite{CKS1,CKS2,CSS1,CSS2}) and to a large extent were studied by
means of numerical simulation (especially because of low
computational cost of the numerical realization of cellular
automata rules, which made it possible to realize large-scale
real-time simulations of urban traffic \cite{SN}).

Roughly speaking the one lane road in these models is associated
to a finite one-dimensional integer lattice of size $N$ with
periodic boundary conditions and each position on the lattice is
either occupied by a particle (represented a vehicle), or empty.
On the next time step each particle remains on its place if the
next position is occupied, and moves forward by one place
otherwise. In \cite{NS,F,BF} it was shown (mainly numerically and
by some physical argument) that limit (as time going to infinity)
behavior of the dynamical system under configuration depends only
on the density of particles in the initial configuration. This
result was generalized for the case of the dynamics on the
infinite lattice and proved mathematically in \cite{Bl-car},
where a novel variational approach was introduced.

Despite various generalizations the one lane restriction of these
models was crucial, for example, in an attempt to study a multi
lane model in \cite{Bl-car}, satisfying standard traffic rules,
no mathematically interesting phenomena were found. Only two
years ago in \cite{NT,NT2} a first nontrivial multi lane
generalization was introduced for the case of a motion on a
finite lattice with periodic boundary conditions, based on a so
called ultradiscrete limit of the well known Burgers equation.

In the present paper is we study ergodic properties of this
model. As we shall show this analysis boils down to the study of
the dynamics of `traffic jams' (see the rigorous definition in
Section~\ref{s:jams}), which mainly depends of the density of
particles in the initial configuration.

One of the main quantity of interest in traffic models -- the
average velocity of cars and its dependence on the density of
cars $\rho$ (called the fundamental diagram) is typically studied
in the steady state. From our results it follows that in the
multi lane model that we consider the average velocity in the
steady state is equal to $\max\{1,K/\rho-1\}$, which immediately
reminds the similar result known in the one lane case.

The paper is organized as follows. In Section~\ref{s:model} we
describe the model in detail and introduce the basic notation
including important notions of dual configurations and maps. In
Section~\ref{s:reg-conf} we introduce the space of regular
(statistically defined) configurations, show that this space is
invariant with respect to dynamics and formulate the main result
of the paper -- Theorem~\ref{t:lim-reg1}. Qualitatively this
result means that in a steady state any configuration either
consists of free (moving independently) particles, or this
property holds for all empty places on the lattice. In terms of
mentioned above variational principle this can be formulated that
the total number of free particles between two fixed ones can
only grow in time. The proof of this result in the next
Section~\ref{s:jams} is based on the detailed analysis of the
dynamics of traffic jams. It is worth notice that in distinction
to the one lane case the formal description of the traffic jam is
rather nontrivial and some individual particles in it can move
still representing obstacles to the motion of other particles.
Section~\ref{s:chaos} is dedicated to the proof of chaoticity of
this model: we explicitly calculate its topological entropy and
show that it is strictly positive. In the last
section~\ref{s:statistics} we derive statistical quantities
describing typical limit configurations.

We tried to define rigorously all important objects that we
consider in the text, however of course we were unable to
introduce all standard mathematical definitions. The reader can
find exact definitions and further references related to
dynamical systems (especially acting on discrete phase spaces),
for example, in \cite{Bl20,ME}.

%%%%%%%%%
\section{Multi lane traffic flow models: dynamics in space of
configurations}\label{s:model}

The model corresponds to the highway traffic flow on a road with
$K$ lanes. Let $\bX_{0}^{K}:=\{0,1,\dots,K\}^{\IZ^1}$ be an
infinite lattice, positions on which we call (lattice) sites. For
a sequence $X\in\bX_{0}^{K}$ and $x\in\IZ^{1}$ by $X(x)$ we
denote the $x$-th element of this sequence.
Consider a map $T:\bX_0^K\to\bX_0^K$, defined by the relation%
\beq{dynamics}{\map X(x) = X(x) + \min\{X(x-1), K-X(x)\}
                                - \min\{X(x),   K-X(x+1)\} .}%

\n{\bf Remark}. In \cite{NT,NT2} the above map were introduced
for the case of the finite lattice with periodic boundary
conditions. Observe that a finite lattice of arbitrary size
$N<\infty$ with periodic boundary conditions is a particular case
of $\IZ^1$ lattice considered in our paper restricted to only
$N$-periodic configurations. The paper \cite{NT2} claims an
estimate of the transient period in the $2N$-periodic case as
$N$. Since the construction in this paper sensitively depends on
the length of the period it cannot be extended to the case of
unbounded lattices with general nonperiodic initial
configurations that we consider. It is of interest that $2N+1$
periodic initial configurations lead (as we shall show) to much
worse estimate of the transient period $2N$.

The above map can be described in a different way in terms of
configurations of particles. Let us introduce several
definitions.

A collection of particles $\Xi$ on the lattice $\IZ^{1}$ we shall
call {\em ordered} if there is a function (called index function)
$I:\Xi\to\IZ^{1}$ such that for any two particles
$\xi,\xi'\in\Xi,~\xi\ne\xi'$ the corresponding indices satisfy
the inequality $I(\xi)\ne I(\xi')$ and if these particles are
located at sites $|\xi|<|\xi'|$ on the lattice then
$I(\xi)<I(\xi')$, where $|\xi|$ stands for the location of the
particle $\xi$ on the lattice.

To a configuration $X\in\bX_{0}^{K}$ we associate an ordered
collection (finite or infinite) of particles on the lattice
$\IZ^{1}$ containing not more than $K$ particles at each site,
such that $X(x)$ for $x\in\IZ^{1}$ means the number of particles
located at the site $x$. Then the set of all possible ordered
configurations of particles containing not more than $K$
particles at each site forms the phase space
$\bX_0^K:=\{0,1,\dots,K\}^{\IZ^1}$ of the system under
configuration.

For a given positive integer $K$ and a given configuration
$X\in\bX_0^K$ the action of the map $T$ can be described as
follows. For each site $x\in\IZ^{1}$ (independently from other
sites) we move $\min\{X(x), K-X(x+1)\}$ particles with the
largest indices from the site $x$ to the site $x+1$.

\begin{lemma} The order function $I$ is preserved under the
action of the map $T$. \end{lemma}

\proof Straightforward. \qed

For a given configuration $X$, associated to the collection of
particles $\Xi$, for each particle $\xi\in\Xi$ we introduce the
notion of velocity $v(\xi)$ which is equal to $1$ if the particle
moves after the application of the map $T$ or $0$ otherwise.
Accordingly we shall say that the particle $\xi$ is {\em free} if
$v(\xi)>0$ and {\em jammed} otherwise. Summing up velocities of
individual particles we obtain moments (the total velocity of
particles at a given site) of lattice sites in the configuration
$X$:
$$ v(X,x) := \sum_{|\xi|=x} v(\xi) .$$
An immediate calculation shows that

\begin{lemma} $v(X,x) = \min\{X(x), \c{X}(x+1)\}$. \end{lemma}

From the point of view of the description in terms of individual
particles we introduce the notion of the dual configuration
$\c{X}(x):=K-X(x)$ for any $x\in\IZ^{1}$, which describes empty
places in the original configurations of particles $X$. Therefore
to describe the dynamics of empty places we consider the {\em
dual map} $\c{\map}:\bX_0^K\to\bX_0^K=\c{(\bX_0^K)}$ whose action
is defined by the relation $\c{\map}X=\c{(\map \c{X})}$ and can
be written as follows.

\begin{lemma} For $X\in\bX_{0}^{K}$ we have
$$ \c{\map} X(x) = X(x) - \min\{X(x), \c{X}(x-1)\}
                        + \min\{X(x+1), \c{X}(x)\} .$$
\end{lemma}

Observe that the dynamics of empty places is exactly the same as
the dynamics of particles, but occurs in the opposite direction.
Obviously both the above formula and the relation
(\ref{dynamics}) describe the mass conservation rule: the number
of particles at a given site in the new configuration is equal to
the number of particles at the same site in the original
configuration minus the number of particles leaving it and plus
the number of particles coming to this site.

By a {\em jammed cluster} (of particles) we shall mean a locally
maximal group of consecutive sites on the lattice containing at
least one jammed particle at each site. Accordingly the jammed
cluster in the dual configuration defines the cluster of free
empty places in the original configuration. The locally maximal
property means that any enlarging of the considered group breaks
the definition, i.e. both immediate neighboring sites to the
cluster do not contain jammed particles.

Consider two subspaces of the space of configurations
$\bX_{0}^{K}$ which shall play an important role in our analysis.
The first of them is the space of configurations of free
particles:
$$ \Free(K) := \{X\in\bX_{0}^{K}:~v(\xi)=1~\forall\xi\in X\} ,$$
and the second one is the space of (space) $n$-periodic
configurations:
$$ \Per(n,K) := \{X\in\bX_{0}^{K}:~X(x)=X(x+n)~\forall x\in\IZ^{1} .\}$$
A trivial calculation show that both of these spaces are
invariant with respect to the dynamics.

\begin{lemma}\label{l:per-inv} $\map:\Free(K)\to\Free(K)$,
$\map:\Per(n,K)\to\Per(n,K)$ for any $n,K\in\IZ^{1}_{+}$.
\end{lemma}

\proof Observe that the restriction of the map $\map$ to the
space of configurations of free particles is equivalent to the
shift operator in this space, from where the first statement
follows immediately. To prove the second statement notice that
according to the formula~(dynamics)
$$ \map X(x) = X(x) + \min\{X(x-1), K-X(x)\}
                    - \min\{X(x),   K-X(x+1)\} \CR
             = X(x+n) + \min\{X(x-1+n), K-X(x+n)\}
                      - \min\{X(x+n),   K-X(x+1+n)\}
             = \map X(x+n) $$
due to the $n$-periodicity of the configuration $X\in\Per(n,K)$.
\qed

Denote by $X=\langle\alpha\rangle$ the $n$-periodic configuration
$X\in\Per(n,K)$ consisting of the periodically repeating word
$\alpha=a_{1}a_{2}\dots a_{n}$ with $a_{i}\in\{0,1,\dots,K\}$ and
such that $X(1)=a_{1}$, for example $X=\langle1234\rangle =
\dots123412341234\dots$.

It is worth notice that despite the statement of the previous
Lemma the minimal period of the configuration may not be
preserved under the dynamics. Indeed, consider a $4$-periodic
configuration $\langle1100\rangle$ and observe that for $K=1$ we
have $\map\langle1100\rangle = \langle1010\rangle \in \Per(2,1)$.

To deal with more general and still statistically homogeneous
configurations in the next section we introduce a more
interesting subset of configurations -- regular configurations
for which as we shall show the statistical description makes
sense.

%%%%
\section{Space of regular configurations}\label{s:reg-conf}

For a configuration $X\in\bX_0^K$ we define the notion of a
subconfiguration $X_{k}^{n}:=\{X(k),X(k+1),\dots,X(n)\}$, i.e. a
collection of entries of $X$ between the pair of given indices
$k<n$, and introduce the corresponding density and the average
velocity:
$$ \rho(X_{k}^{n}) := \frac{m(X_{k}^{n})}{n-k+1}, \qquad
     \v(X_{k}^{n}) := \frac1{m(X_{k}^{n-1})} \sum_{x=k}^{n-1}v(X,x) ,$$
where $m(X_{k}^{n}):=\sum_{x=k}^{n}X(x)$ stays for the number of
particles in the subconfiguration $X_{k}^{n}$. \footnote{Observe
that in the definition of the average velocity we consider only
particles from sites till $n-1$. This is related to the fact that
velocities of particles in the site $n$ is not defined by the
subconfiguration $X_{k}^{n}$.}

By the {\em density} and the {\em average velocity} (of
particles) of a entire configuration $X\in\bX_0^K$ we mean the
following limits (if they are well defined):
$$ \rho(X) := \lim_{n\to\infty}\rho(X_{-n}^{n}), \qquad
     \v(X) := \lim_{n\to\infty}\v(X_{-n}^{n}) ,$$
otherwise one can consider the corresponding upper and lower
limits, which we denote by $\rho_{\pm}(X)$ and $\v_{\pm}(X)$.

Notice that both these quantities are well defined in the case of
space periodic configurations (belonging to $\Per(K)$) in
distinction even to the simplest case when a configuration $X$
consists of free particles (i.e. belongs to $\Free(K)$). Thus in
the general case important statistical quantities $\rho(X),
\v(X)$ may be not well defined. To be able to deal with the space
of configurations satisfying a reasonable statistical description
we introduce the following space of configurations.

We shall say that a configuration $X$ satisfies the {\em
regularity} assumption (or simply {\em regular}) if there exists
a number $\rho\in[0,K]$ and a strictly monotone function $\phi(n)
\to 0$ as $n\to\infty$ (which we call rate function), such that
for any $n\in\IZ^{1}$, $N\in\IZ_{+}^{1}$ and any subconfiguration
$X_{n+1}^{n+N}$ of length $N$ the number of particles in this
subconfiguration
$m(X_{n+1}^{n+N})$ satisfies the inequality %
\beq{reg-ineq}{
 \left| \frac{m(X_{n+1}^{n+N})}N - \rho \right| \le \phi(N) .}%
It is clear that at for a configuration $X$ satisfying the
regularity assumption the density of particles $\rho(X)$ is well
defined and is equal to the value $\rho$ in the formulation of
the assumption. The space of configurations from $X_{0}^{K}$
satisfying the regular assumption with the density $\rho$ and the
rate function $\phi$ we shall denote by $\Reg(\rho,\phi,K)$.

The main result of the paper formulated below describes the
restriction of the dynamics to the space of regular
configurations and will be proven in the rest of this section and
the next one.

\begin{theorem}\label{t:lim-reg1} Let the initial configuration
$X \in\ Reg(\rho,\phi)$ with $\rho\ne K/2$. Then after a finite
number of iterations $t\le t_c=t_c(\rho,\phi) := \frac14
(\phi^{-1}(\frac{K}2-\rho)+1)^2$ for the configuration $T^tX$ the
average velocity of particles becomes well defined and is equal
to $\min\{1, \frac{K}\rho-1\}$. Moreover for any $t\ge t_c$ we
have $T^tX\in\Free(K)$ if $\rho<K/2$ and $\c{(T^tX)}\in\Free(K)$
if $\rho>K/2$.
\end{theorem}

To analyze properties of regular configurations we introduce the
binary relation `domination', which we denote by $\rel$, on the
set of configurations $\bX_0^K$ as follows: $X\rel Y$ if and only
if for any $n\in\IZ, N\in\IZ^{+}$ there exists a pair
$n_{-},n_{+}\in\IZ$ such that
$$ m(X_{n_{-}+1}^{n_{-}+N}) \le m(Y_{n+1}^{n+N})
                            \le m(X_{n_{+}+1}^{n_{+}+N}) .$$

\begin{lemma} The relation $\rel$ is an order relation, i.e. it
is reflexive and transitive, but this relation is not symmetric.
\end{lemma}

\proof The proof of the reflexivity, i.e. that $X\rel X$ for any
$X\in\bX_0^K$ is straightforward. To prove the second statement
consider a pair of configurations $X\rel Y \rel Z$. By definition
for any $n,k\in\IZ$, $N\in\IZ^{+}$ we have
$$ m(X_{n_{-}+1}^{n_{-}+N}) \le m(Y_{n+1}^{n+N})
                            \le m(X_{n_{+}+1}^{n_{+}+N}), $$
$$ m(Y_{n'_{-}+1}^{n'_{-}+N}) \le m(Z_{k+1}^{k+N})
                            \le m(Y_{n'_{+}+1}^{n'_{+}+N}) .$$
Thus for any $k,N$ there exists $n_{-},n_{+},n'_{-},n'_{+}\in\IZ$
such that
$$ m(X_{n_{-}+1}^{n_{-}+N}) \le m(Y_{n'_{-}+1}^{n+N})
      \le m(Z_{k+1}^{k+N})  \le m(Y_{n'_{+}+1}^{n'_{+}+N})
                            \le m(X_{n_{+}+1}^{n_{+}+N}) .$$
Therefore $X\rel Z$. It remains to prove the absence of symmetry,
i.e. that there exists a pair of configurations $X\rel Y$ such
that the relation $Y\rel X$ does not hold. Let $X(1)=1$, while
$X(x)=0$ for all $x\ne1$, and let $Y(x)=0$ for all $x\in\IZ$.
Then clearly $X\rel Y$ but the opposite relation does not hold.
\qed

\begin{lemma}\label{l:rel} Let $X\rel Y$ and let $X\in\Reg(\rho,\phi,K)$.
Then $Y\in\Reg(\rho,\phi,K)$ as well.
\end{lemma}

\proof According to the definition for any $n,N$ there exists a
pair $n_{-},n_{+}$ such that
$$ m(X_{n_{-}+1}^{n_{-}+N}) \le m(Y_{n+1}^{n+N})
                            \le m(X_{n_{+}+1}^{n_{+}+N}) .$$
Thus
$$ -\phi(N) \le \frac{m(X_{n_{-}+1}^{n_{-}+N})}N - \rho
            \le \frac{m(Y_{n+1}^{n+N})}N - \rho
            \le \frac{m(X_{n_{+}+1}^{n_{-}+N})}N - \rho \le \phi(N),$$
which yields the desired statement. \qed

Now we are ready to prove that the set of regular configurations
is invariant under the dynamics.

\begin{lemma}\label{l:reg-inv}
$\map:\Reg(\rho,\phi,K)\to \Reg(\rho,\phi,K)$ for any triple
$(\rho,\phi,K)$. \end{lemma}

\proof Let a configuration $X\in\Reg(\rho,\phi,K)$. We need to
show that the configuration $\map X$ also satisfies the same
assumption. To do it we shall prove that $X\rel\map X$, from
where by Lemma~\ref{l:rel} we shall get the desired statement.
Fix arbitrary integers $n\in\IZ$ and $N\in\IZ^{+}$ and consider
the subconfiguration $(\map X)_{n+1}^{n+N}$. The number of
particles in this subconfiguration differs from the number of
particles in the subconfiguration $X_{n+1}^{n+N}$ by the number
of particles $P_{-}$ coming from the site $n$ to the site $n+1$
and the number of particles $P_{+}$ coming from the site $n+N$ to
the site $n+N+1$, i.e.
$$ m((\map X)_{n+1}^{n+N}) = m(X_{n+1}^{n+N}) + P_{-} - P_{+} .$$
There might be four possible situations:

(a) $X(n)+X(n+1)\le K$ and $X(n+N)+X(n+N+1)\le K$. Then
$P_-=X(n)$, $P_+=X(n+N)$, and thus
$$ m((\map X)_{n+1}^{n+N}) = m(X_{n+1}^{n+N}) + X(n) - X(n+N)
 = m(X_{n}^{n+N-1}) .$$

(b) $X(n)+X(n+1)\le K$ and $X(n+N)+X(n+N+1) > K$. Then
$P_-=X(n)$, $P_+=K-X(n+N+1)$, and
$$ m((\map X)_{n+1}^{n+N}) = m(X_{n+1}^{n+N}) + X(n) + K - X(n+N+1) \CR
 = m(X_{n}^{n+N-1}) + X(n+N) + X(n+N+1) - K
 > m(X_{n}^{n+N-1}) .$$
On the other hand,
$$ m((\map X)_{n+1}^{n+N}) = m(X_{n+1}^{n+N}) + X(n) + K - X(n+N+1) \CR
 = m(X_{n+2}^{n+N+1}) + X(n) + X(n+1) - K
 \le m(X_{n+2}^{n+N+1}) .$$

(c) $X(n)+X(n+1) > K$ and $X(n+N)+X(n+N+1) \le K$. Then
$P_-=K-X(n+1)$, $P_+=X(n+N)$, and
$$ m((\map X)_{n+1}^{n+N}) = m(X_{n+1}^{n+N}) + K - X(n+1) - X(n+N) \CR
 = m(X_{n}^{n+N-1}) - X(n) + X(n+N) + K - X(n+1) - X(n+N)
 > m(X_{n}^{n+N-1}) .$$
On the other hand,
$$ m((\map X)_{n+1}^{n+N}) = m(X_{n+1}^{n+N}) + K - X(n+1) - X(n+N) \CR
 = m(X_{n+2}^{n+N+1}) - X(n+N+1) + K - X(n+N)
 \le m(X_{n+2}^{n+N+1}) .$$

(d) $X(n)+X(n+1) > K$ and $X(n+N)+X(n+N+1) > K$. Then $P_-=K-X(n+1)$,
$P_+=K-X(n+N+1)$ and
$$ m((\map X)_{n+1}^{n+N}) = m(X_{n+1}^{n+N}) + K - X(n+1) - K + X(n+N+1)
 = m(X_{n+2}^{n+N+1}) .$$

Therefore in all four possible cases we have found
subconfigurations in $X$ approximating (by the number of
particles) those in $\map X$ from both hands, which yields the
statement of Lemma. \qed

\begin{lemma}\label{l:dual-reg}
$\c{(\Reg(\rho,\phi,K))}=\Reg(K-\rho,\phi,K)$.
\end{lemma}

\proof Let $X\in\Reg(\rho,\phi,K)$. Then for any $n\in\IZ^{1},
N\in\IZ^{1}_{+}$ we have
$$ m((\c{X})_{n+1}^{n+N})
 = m((\langle K\rangle - X)_{n+1}^{n+N})
 = K\cdot N - m(X_{n+1}^{n+N}) .$$
Therefore
$$ \left|\rho((\c{X})_{n+1}^{n+N}) - (K-\rho)\right|
 = \left|\frac{m((\c{X})_{n+1}^{n+N})}{N} - (K-\rho)\right| \CR
 = \left|\frac{m(X_{n+1}^{n+N})}{N} - \rho)\right| \le \phi(N) .$$
\qed

Consider now the connection between spaces of periodic
configurations and regular ones. Clearly, for any configuration
$X\in\Per(n,K)$ the notion of density $\rho(X)$ is well defined
and $\rho(X)=m(X_{1}^{n})/n$. To specify the density we denote by
$\Per_{\rho}(n,K)$ the set of configurations from $\Per(n,K)$
having the same density $\rho$.

\begin{lemma} For any $\rho,n,K$ we have
$\map:\Per_{\rho}(n,K)\to\Per_{\rho}(n,K)$ and
$\Per_{\rho}(n,K)\subset\Reg(\rho,~\rho(1-\frac{\rho}{K})\frac{n}{N},~K)$.
\end{lemma}

\proof For a given configuration $X\in\Per(n,K)$ denote
$\rho:=\rho(X)=m(X_{1}^{n})/n$. The first statement immediately
follows from Lemma~\ref{l:per-inv} and the fact that the number
of particles on the period of the configuration cannot change
under dynamics. Now each positive integer $N$ can be represented
as $N=kn+l$, where $k\in\{0,1,\dots\}$, $l\in{0,1,\dots,n-1}$.
For any $l\le n-\frac{\rho}{K}n$ the number of particles in the
subconfiguration $X_{x+1}^{x+N}$ can be estimated from below as
$$ m(X_{x+1}^{x+N}) \ge \rho kn .$$
Therefore
$$ \rho - \frac{m(X_{x+1}^{x+N})}{N}
 \le \rho - \frac{\rho kn}{kn+l} = \frac{\rho l}{N}
 \le \rho(1-\frac\rho{K})\frac{n}{N} =: \phi(N) .$$
Otherwise, if $l>n-\frac{\rho}{K}n$ we have
$$ m(X_{x+1}^{x+N}) \ge \rho kn + K(l-n+\frac{\rho}{K}n)
                      = \rho kn + Kl - Kn + \rho n .$$
Thus
$$ \rho - \frac{m(X_{x+1}^{x+N})}{N}
 \le \rho - \frac{\rho kn + Kl - Kn + \rho n}{kn+l} \CR
 = \frac1N (n-l)(K-\rho) < \frac1N \frac{\rho n}{K}(K-\rho)
 = \phi(N) .$$
Now we shall use estimates for the number of particles in the
subconfiguration $X_{x+1}^{x+N}$ from above. If
$l\le\frac\rho{K}n$ then
$$ m(X_{x+1}^{x+N}) \le \rho kn + Kl $$
and
$$ \frac{m(X_{x+1}^{x+N})}{N} - \rho
 \le \frac{\rho kn + Kl}{kn+l} - \rho
   = \frac1N (K-\rho)l \le \frac1N (K-\rho)\frac\rho{K}n = \phi(N) .$$
Otherwise, if $l>\frac\rho{K}n$ then
$$ m(X_{x+1}^{x+N}) \le \rho kn + \rho n $$
and
$$ \frac{m(X_{x+1}^{x+N})}{N} - \rho
 \le \frac{\rho kn + \rho n}{kn+l} - \rho
   = \frac1N \rho(n-l) \CR
   < \frac1N \rho(n-\frac\rho{K}n)
   = \frac1N \rho n(1-\frac\rho{K}) = \phi(N) .$$
\qed

One can easily check that for any (space) periodic configuration
the notion of the average velocity is well defined. It is of
interest that for more general class of regular configurations
this is not the case even for $K=1$. Denote $a=1100$ and $b=1010$
and consider the configuration $X$ constructed as follows: %
\beq{reg-non-velocity}{\dots~bbbbaaaa~bbaa~ba~ab~aabb~aaaabbbb~\dots ,}%
i.e. $X_{1}^{8}=ab$, $X_{-7}^{0}=ba$, $X_{9}^{24}=aabb$,
$X_{-23}^{-8}=bbaa$, etc. Notice that in each subsequent series
the number of consequent elements $aa\dots a$ and $bb\dots b$
doubles.

\begin{lemma} The configuration $X$ defined as
(\ref{reg-non-velocity}) is regular ($X\in\Reg(1/2,1/N,K)$), but
the average velocity is not well defined. \end{lemma}

\proof Observe that $m(X_{i+1}^{i+4k})=2k$ for any $i,k$, while
$2k \le m(X_{i+1}^{i+4k+j}) \le 2k+2$ for any $j\in\{1,2,3\}$.
Therefore the configuration (\ref{reg-non-velocity}) is regular
with the density $1/2$ and the rate function $\phi(N)=1/N$.

Let us calculate now the average velocity on various
subconfigurations. First consider a subconfiguration starting
from the 1st element and containing the full series $aa\dots bb$,
i.e. $X_{1}^{2(2^{k+1}-1)}$. This subconfiguration for any $k$
contains the same number of elements $a$ and $b$ and hence
$$ \v(X_{1}^{2(2^{k+1}-1)})
 = \frac12 \cdot\frac12 + \frac12 \cdot1 = \frac34 .$$
Similarly due to the symmetry of the configuration $X$ we have
$\v(X_{-2(2^{k+1}-1)+1}^{0})=3/4$ and therefore
$|\v(X_{-2^{k+2}}^{2^{k+2}}) - 3/4| < 5\cdot 2^{-(k+3)}$. Thus
$$ \v(X_{-2^{k+2}}^{2^{k+2}}) \to 3/4 \quad {\rm as} \quad k\to\infty .$$

Another type of subconfigurations that we consider differs from
the previous one by the fact that it contains an additional
(full) series of elements $a\dots a$ in the end, i.e.
$X_{1}^{2(2^{k+1}-1)+2^{k+1}}= X_{1}^{3\cdot2^{k+1}-2}$.
$$ \v(X_{1}^{3\cdot2^{k+1}-2}) =
   \frac{\frac12 \cdot 2^{k+1} + \frac34 \cdot 2(2^{k+1}-1)}
        {2^{k+1} + 2(2^{k+1}-1)}
 = \frac{2\cdot 2^{k+1} - \frac32}{3\cdot2^{k+1} -2}
 \to \frac23 $$
as $k\to\infty$. Therefore using again the symmetry of the
configuration $X$ we have
$$ \v(X_{-3\cdot2^{k+1}}^{3\cdot2^{k+1}}) \to 2/3
                           \quad {\rm as} \quad k\to\infty ,$$
and thus different subsequences of $k$ lead to different average
velocities. \qed

%%%%
\section{Traffic jams and simple properties of the
dynamics}\label{s:jams}

Recall that in a configuration $X$ sites between $x'$ and $x''$
belong to the {\em jammed cluster} if for any integer
$x\in[x',x'']$ the inequality $X(x)+X(x+1)>K$ holds true.
Similarly consecutive sites for which this inequality does not
hold belong to a free cluster. The site $x$ is called {\em free}
if $X(x)+X(x+1)\le K$.

\begin{lemma}\label{l:min-max}
For each configuration $X \in \bX_0^K$ and for each site
$x\in\IZ^{1}$ we have
$$ \min\{X(x-1), X(x), X(x+1)\} \le (TX)(x)
  \le \max\{X(x-1), X(x), X(x+1)\} $$
and thus for any $t\in\IZ^{1}_{+}$
$$ \min_{x}\{X(x)\} \le \min_{x}\{(T^{t}X)(x)\}
   \le \max_{x}\{(T^{t}X)(x)\} \le \max_{x}\{X(x)\} .$$
Moreover, the upper and lower limits may be not preserved under
dynamics:

-- $\exists X\in\bX_0^K$ such that
   $\max_{x}\{(TX)(x)\} < \max_{x}\{X(x)\}$;

-- $\exists X\in\bX_0^K$ such that
   $\min_{x}\{(TX)(x)\} > \min_{x}\{X(x)\}$;
\end{lemma}

\proof Fix a configuration $X\in\bX_0^K$ and a site
$x\in\IZ^{1}$. Then denoting by $P(x):=\min\{X(x-1), K-X(x)\}$
the number of particles moving to the site $x$ (from the site
$x-1)$) under the action of the map $\map$ we get
$$ (TX)(x) = X(x) + P(x) - P(x+1) .$$
Consider all 4 possibilities: \par\n%
(a) $X(x-1)\le K-X(x)$ and $X(x)\le K-X(x+1)$. Thus
    $$ (TX)(x) = X(x) + X(x-1) - X(x) = X(x-1) .$$
(b) $X(x-1) > K-X(x)$ and $X(x)\le K-X(x+1)$. Thus
    $$ (TX)(x) = X(x) + (K-X(x)) - X(x) = K - X(x) < X(x-1) .$$
    \qquad On the other hand, in this case
    $$ (TX)(x) = K - X(x) \ge X(x+1) .$$
(c) $X(x-1)\le K-X(x)$ and $X(x) > K-X(x+1)$. Thus
    $$ (TX)(x) = X(x) + X(x-1) - (K-X(x+1))
     = X(x-1) - (K-X(x)) + X(x+1) \le X(x+1) .$$
    \qquad On the other hand, in this case
    $$ (TX)(x) = X(x-1) + X(x) - (K-X(x+1))
    > X(x-1) + X(x) - X(x) = X(x-1) .$$
(d) $X(x-1) > K-X(x)$ and $X(x) > K-X(x+1)$. Thus
    $$ (TX)(x) = X(x) + (K-X(x)) - (K-X(x+1)) = X(x+1) .$$
Thus the first statement of Lemma holds true in all situations.

It remains to construct examples of configurations satisfying the
last two statements of Lemma. Let $K=2$. Then
$T:\langle221022\rangle \to \langle211122\rangle$ and
$T:\langle002100\rangle \to \langle001110\rangle$. In the first
example the minimal value $0$ becomes $1$, while in the second
example the maximal value $2$ becomes $1$ under the action of the
dynamics. \qed

Introduce a map marking global maxima of a configuration
$M:\bX_{0}^{K}\to\bX_{0}^{1}$ as follows: $MX(x):=1$ if
$X(x)=\max_{y}\{X(y)\}$ and $MX(x):=0$ otherwise. We define also
arithmetic operations with configurations $X,Y\in\bX_{0}^{K}$:
$$ (X+Y)(x) := \min\{X(x)+Y(x),~K\}, \qquad
   (X-Y)(x) := \max\{X(x)-Y(x),~0\}. $$
Using this notation we can formulate the following decomposition
result.

\begin{lemma}\label{l:decomp} For a given $X\in\bX_{1}^{K}$ if
$\forall x\in\IZ^{1}_{+}$ such that $MX(x)=1$ holds
$X(x-1)+X(x)\le K$ and $X(x)+X(x+1)\le K$ then
$$ \map X = \map(X-MX) + \map(MX) ,$$
otherwise
$$ \map X = \map_{|_{K-1}}(X-MX) + \map_{|_{1}}MX ,$$
where $\map_{|_{K-1}}$ means the restriction of the map $\map$ to
$\bX_{0}^{K-1}$. On the other hand, there exists a configuration
$X\in\bX_0^K$ such that
$$ (\map^{t}X)(x) \ne (\map_{|_{K-1}}^{t}X)(x) .$$
even if $\max_{x}\{X(x)\}<K$.
\end{lemma}

\proof The statement about the decomposition follows immediately
from the definition of the dynamics while the following example
demonstrate the second statement:
$\map_{|_{3}}: \langle1221\rangle \to \langle1212\rangle, \quad
 \map_{|_{2}}: \langle1221\rangle \to \langle2211\rangle$. \qed

These results demonstrate rather counter intuitive properties of
the considered model of traffic flows. For example, from
Lemma~\ref{l:min-max} it follows that if for a given initial
configuration one traffic lane is not occupied (along the entire
lattice), then these property holds for any moment of time.
So it looks that the dynamics will not change if the road will be
made narrower by one lane. However this is completely wrong, which
was demontstrated in the second statement of Lemma~\ref{l:decomp}.

Another example gives the following seemingly evident (but wrong)
decomposition, which one would expect instead of the more complex
decomposition described in Lemma~\ref{l:decomp}. Assume that for
a configuration $X$ we have $X(x)>1$ for all $x\in\IZ^1$. Then it
looks reasonable that the dynamics of the configuration, restricted
to the lanes $2,3,\dots,K$ should be the same as in the original one,
i.e.
$$  \map_{|_{K}} X = \map_{|_{K-1}}(X - \langle1\rangle)
                   + \map_{|_{K}}(\langle1\rangle) .$$
The following example of a periodic configuration shows that this
is not the case:
$$  \map_{|_2}\langle1221\rangle = \langle2211\rangle, \qquad
    \map_{|_1}\langle0110\rangle + \map_{|_2}\langle1111\rangle
          = \langle0101\rangle + \langle1111\rangle
          = \langle1212\rangle .$$

\begin{lemma}\label{cluster-repr}
Let $X_{k+1}^{k+n}$ be a jammed cluster of length $n$ in the
configuration $X$. Then %
\beq{eq:cl-1}{ (\map X)(x) = X(x+1) \quad \forall x\in\{k+2,\dots,k+n\} ,}%
\beq{eq:cl-2}{ (\map X)(k+n+1) = K - X(k+n+1) ,}%
\beq{eq:cl-3}{ (\map X)(k+1) = X(k) + X(k+1) + X(k+2) - K }%
and if the site $k-1$ does not belong to another jammed cluster, then
\beq{eq:cl-4}{ (\map X)(k-1) = X(k-2), \quad (\map X)(k) = X(k-1) ,}%
otherwise %
\beq{eq:cl-5}{ (\map X)(k-1) = X(k), \quad (\map X)(k) = K - X(k) .}%
\end{lemma}

\proof First let us show that $\map X(x)=X(x+1)$ for all
$x\in\{k+2,\dots,k+n\}$. Observe that by the definition of a jammed
cluster we have
$$ X(x-1)+X(x)>K, \quad X(x)+X(x+1)>K .$$
Thus
$$ X(x-1)>K-X(x), \quad X(x)>K-X(x+1) .$$
Therefore after the application of the map $\map$ exactly $K-X(x)$
particles comes to the site $x$ from the site $x-1$, while $K-X(x+1)$
particles leaves it. Therefore
$$ \map X(x) = K-X(x) + X(x) - (K-X(x+1)) = X(x+1) ,$$
which proves the equality~(\ref{eq:cl-1}). Observe that this equality
makes sense only if $n\ge 2$.

The site $k+n+1$ is the first free site after the jammed cluster. Therefore
all particles from it moves to the site $k+n+2$ under the action of the
map $\map$, while exactly $K - X(k+n+1)$ particles moves to the site
$k+n+1$
from the last site of the considered jammed cluster. This gives the
equality~(\ref{eq:cl-2}). Notice that from this inequality we get that
$$ (\map X)(k+n) + (\map X)(k+n+1) = K ,$$
i.e. the site $k+n$ does not belong to a jammed cluster in the
configuration $TX$.

Clearly the site $k$ cannot belong to another jammed cluster, otherwise
the site $k+1$ would not be the first site of the considered jammed
cluster.

By definition under the action of the map $\map$ all particles from the
site $k$ moves to the site $k+1$, from where exactly $X(k+2) - K$ particles
moves to the site $k+2$. Thus we get the equality~(\ref{eq:cl-3}).

Consider now the case when the site $k-1$ does not belong to another jammed
cluster, i.e. $X(k-1)+X(k)\le K$. This immediately gives the
formulae~(\ref{eq:cl-4}) for the number of particles in the sites $k-1$
and $k$.

If $n=1$ then
$$ (TX)(k+1) + (TX)(k+2) = X(k) + X(k+1) + X(k+2) - K + K - X(k+2)
                         = X(k) + X(k+1) \le K .$$
If $n\ge2$ then
$$ (TX)(k+i) + (TX)(k+i+1) = X(k+i-1) + X(k+i) $$
for all $i\in\{1,\dots,n-2\}$. Thus
$$ (TX)(k+n-1) + (TX)(k+n) = X(k+n) + X(k+n+1) > K , \CR
   (TX)(k+n) + (TX)(k+n+1) = X(k+n+1) + K - X(k+n+1) = K  .$$
Therefore in both cases the site which was the last one in the jammed
cluster $X_{k+1}^{k+n+1}$ becomes, and if $n\ge2$ the site $k+n-1$ turns
out
to be the last site in the jammed cluster in $TX$.

Summarizing, in the case when the site $k-1$ is free we get the
following representation for $\map X$ in the neighborhood of the
considered jammed cluster:%
{\small\begin{verbatim}
    ...   k-1     k       k+1       k+2     k+3   ...  k+n-1    k+n     
k+n+1     ...
 X= ...  X(k-1)  X(k)   [X(k+1)    X(k+2)  X(k+3) ... X(k+n-1) X(k+n)] 
X(k+n+1)   ...
TX= ...  X(k-2)  X(k-1) (TX)(k+1)  X(k+3)  X(k+4) ... X(k+n)]  X(k+n+1)
K-X(k+n+1) ...
\end{verbatim}}
By square brackets we denote the boundaries of jammed clusters. Observe
that
the first site of the new cluster might be either $k$ or $k+1$.

In the alternative case when the site $k-1$ is the last site of
the previous jammed cluster, the representation differs only at
sites $k-1$ and $k$:%
{\small\begin{verbatim}
    ...   k-1     k       k+1       k+2     k+3   ...  k+n-1    k+n     
k+n+1     ...
 X= ...  X(k-1)]  X(k)   [X(k+1)   X(k+2)  X(k+3) ... X(k+n-1) X(k+n)] 
X(k+n+1)   ...
TX= ...  X(k)    [K-X(k) (TX)(k+1) X(k+3)  X(k+4) ... X(k+n)]  X(k+n+1)
K-X(k+n+1) ...
\end{verbatim}}
Indeed, applying the first statement of Lemma (which has been already
proven) to the previous cluster, we get that $(\map X)(k-1)= X(k)$.
On the other hand, the number of particles moving from the site $k-1$ to
the site $k$ is equal to $K-X(k)$, while all the particles that were
at site $k$ move to the site $k+1$ (since the site $k$ does not belong
to a jammed cluster).
$$ (TX)(k) + (TX)(k+1) = K - X(k) + X(k) + X(k+1) + X(k+2) - K  \CR
                       = X(k+1) + X(k+2) > K ,$$
therefore the new jammed cluster has the same length and
is located at the sites from $k$ to $k+n-1$.
\qed

\begin{corollary}\label{col:cluster1} For each jammed cluster
$(TX)_{k+1}^{k+n+1}$ of length $n>1$ we have
$(TX)(k+n+2) + (TX)(k+n+3) = K$.  \end{corollary}

\proof Immediately follows from the equality~(\ref{eq:cl-2}). \qed

This implies that the distance between two consecitive jammed clusters
is at least 2.

\begin{lemma}\label{cluster-size}
Let $X_{k+1}^{k+n}$ be a jammed cluster of length $n$ in the configuration
$X$. Then neither its length, nor the number of particles in it cannot
increase under dynamics. Moreover, if $X(k-1) + X(k)<K$ then number of
particles in the jammed cluster decreases at least by $K-(X(k-1) + X(k))>0$
after the application of the map.
\end{lemma}

\proof Consider first the case when the site $k-1$ does not belong to
another jammed cluster, i.e.
$$ X(k-1) + X(k) \le K, \qquad X(k) + X(k+1) \le K .$$
Clearly, in this case the site $k-1$ cannot be the first site of
the jammed cluster in the configuration $\map X$. Therefore in
the worst case the jammed cluster is located at sites from $k$ to
$k+n-1$, i.e. its length is at least not larger than of the
considered one. Applying Lemma~\ref{cluster-repr} we can estimate
from above the difference between the number of particles in the
new cluster and the old one as follows:
$$ [(\map X)(k) + (\map X)(k+1)\} - \{X(k+1) + X(k+2)] \CR
 = X(k-1) + X(k) + X(k+1) + X(k+2) - K - X(k+1) - X(k+2) \CR
 = X(k-1) + X(k) - K \le 0 .$$
Hence the number of particles in this case cannot increase, and moreover
this number decreases if $X(k-1) + X(k)<K$.

It remains to consider the case when the considered jammed cluster is
immediately preceded by another jammed cluster. Again by
Lemma~\ref{cluster-repr}
$$ (\map X)(k-1) + (\map X)(k) = X(k) + K - X(k) = K .$$
Hence the site $k-1$ does not belong to the jammed cluster. On the other
hand,
$$ (\map X)(k) + (\map X)(k+1)
 = K - X(k) + X(k) + X(k+1) + X(k+2) - K  \CR
 = X(k+1) + X(k+2) > K ,$$
since the site $k+1$ belongs to the jammed cluster. Thus the site $k+1$
is the first site of the jammed cluster in the configuration $\map X$,
which lies in sites from $k$ to $k+n$, i.e. its length is exactly the
same as of the old one. Applying the same trick as above to calculate
the difference between the number of particles in the new cluster and
the old one we get:
$$ \{(\map X)(k) + (\map X)(k+1)\} - \{X(k+1) + X(k+2)\} \CR
 = K - X(k) + X(k) + X(k+1) + X(k+2) - K - X(k+1) - X(k+2) = 0 .$$
Therefore even the number of particles in the jammed cluster is preserved
in this case. \qed

\begin{lemma}\label{cluster-dens}
Let $n\in\IZ_{+}^{1}$, $\rho(X_{k+1}^{k+2n+1})\le K/2$, and let
in the subconfiguration $X_{k+1}^{k+2n+1}$ there is at least one
jammed site. Then there is an integer $i\in\{1,\dots,2n-1\}$ such
that $X(k-i) + X(k+i+1)<K$.
\end{lemma}

\proof Assume that this statement does not hold true. Then for
any two consecutive sites $x$ and $x+1$ in this subconfiguration
we have $X(x) + X(x+1)\ge K$. On the other hand, for the jammed
site $y$ we get $X(x) + X(x+1)\ge K+1$. Thus
$$ \sum_{x=1}^{2n}X(k+x) \ge n+1 ,$$
which contradicts to the fact that the density is less or equal to $1/2$.
\qed

These results yield the following property: For any given
subconfiguration the number of particles in any jammed cluster
completely contained in this subconfiguration is a nonincreasing
function of time and achieves its lowest possible level under
dynamics.

%\bigskip
%Examples:\quad%
%$\map_{|_{4}}(\dots0[4142]30\dots) = \dots0[142]313\dots$,
%$\map_{|_{3}}(\dots0[3132]20\dots) = \dots0[132]212\dots$,
%$\map_{|_{5}}(\dots1132[3434]210\dots) = \dots11[3434]232\dots$.

\begin{figure} \begin{center}
\begin{verbatim}
        t     X     v(X)        X     v(X)
        0  <21200>  3/5      <22300>  4/7
        1  <12020>  4/5      <13030>  6/7
        2  <10202>  4/5      <10303>  6/7
        3  <11021>  4/5      <21031>  6/7
        4  <11111>   1       <12112>   1
             (a)               (b)
\end{verbatim}
\end{center} \caption{Two examples of the dynamics of $n$-periodic
configurations:  (a) $K=2$, $n=5$, $\rho=1=K/2$, \ \
(b) $K=3$, $n=5$, $\rho=7/5<K/2=3/2$. \label{ex:1}}
\end{figure}

\begin{lemma}\label{l:tr-per-reg} Let $X\in\Reg(\rho,\phi,K)$ with
the density $\rho<K/2$. Then after at most
$t_c=t_c(\rho,\phi)=\frac14(\phi^{-1}(\frac{K}2-\rho)+1)^2$
iterations all particles in $T^{t}X$ for $t\ge t_{c}$ will become
free.
\end{lemma}

\proof According to the definition of regular configurations
$M(n)$ -- the maximal number of particles in subconfigurations of
length $n$ of the configuration $X$ for each $n\in\IZ^{1}_{+}$
satisfies the inequality
$$ \frac{M(n)}n \le \rho + \phi(n) .$$
Thus for any $n>N_c:=\phi^{-1}(\frac{K}2-\rho)$ it follows that
$M(n)<n/(2K)$. By Lemma~\ref{cluster-dens} in each
subconfiguration of length $n$ there is a pair of consequent
sites whose total number of particles $Q$ is strictly less than
$K$. Consider the dynamics of this pair of sites. According to
our previous results, while the site ahead of them is free these
two sites will simply move one position forward. In the opposite
case, when the next site is the first site of some jammed cluster
by Lemma~\ref{cluster-size} the number of particles in this
cluster will decrease by $K-Q>0$. On the other hand, free
particles and jammed clusters move in opposite directions each
with the velocity $1$. Thus the maximum time between the
consecutive meetings of a jammed cluster and a pair of consequent
sites with the total number of particles less than $K$ does not
exceed $n/2$. Let $n$ be the smallest integer larger or equal to
$N_{c}$. Since after each such meeting the number of particles in
the corresponding jammed cluster decreases at least by 1 and
since the number of particles in this cluster is less or equal to
$M(n)$ we get the following upper estimate of the transient
period:
$$ t_{c} \le \frac{M(n)\cdot n}2 \le \frac{n^{2}}{4K}
         < \frac14 (\phi^{-1}(K/2-\rho)+1)^2 .$$
\qed

\n{\bf Proof} of Theorem~\ref{t:lim-reg1}. After the preparation
made in Lemmas~\ref{l:rel}--\ref{l:tr-per-reg} we are able to
finish the proof of our main result. Indeed, in the case of a
regular configuration $X\in\Reg(\rho,\phi,K)$ with the density
$\rho<K/2$ by Lemma~\ref{l:tr-per-reg} for any integer $t\ge
t_{c}$ the configuration $\map^{t}X$ consists of only free
particles. In the opposite case, when $\rho>K/2$, we consider the
dual configuration $\c{X}\in\Reg(K-\rho,\phi,K)$ (by
Lemma~\ref{l:dual-reg}) and the since the action of the dual map
is equivalent to the main one but proceeds in the opposite
direction we get that $\c{\map^{t}X}\in\Free(K)$.

It remains to prove the statement about the average velocity of
the configuration $\map^{t}X$ for each $t\ge t_{c}$. Again we
start from the case of low density $\rho<K/2$. Since the
configuration $\map^{t}X$ consists of free particles, velocity of
each particle is equal to $1$. Thus
$$ V((\map^{t}X)_{-n}^{n}) \equiv 1 \quad \forall n\in\IZ^{1}_{+},$$
which both shows that the average velocity s well defined and
that $\v(\map^{t}X)=1$. Now if $\rho>K/2$ the dual configuration
to the configuration $Y:=\map^{t}X$ again consists of free
particles. Hence
$$ V(Y_{-n}^{n}) = \frac1{m(Y_{-n}^{n-1})} \sum_{x=-n}^{n-1}v(Y,x)
 = \frac{m((\c{Y})_{-n}^{n-1})}{2nK - m((\c{Y})_{-n}^{n-1})}
 \to \frac{K}\rho -1 $$
as $n\to\infty$ since the density of the configuration $\c{Y}$ is
equal to $K-\rho$. \qed

Observe that in the proof of Theorem~\ref{t:lim-reg1} we actually
derived an estimate of the length of the transient period as
$t\le t_c=t_c(\rho,\phi) := \frac14
(\phi^{-1}(\frac{K}2-\rho)+1)^2$ which goes to infinity as
$\rho\to1/2$. This is the reason why Theorem~\ref{t:lim-reg1}
does not cover the boundary case $\rho=1/2$, which we discuss
below.

\begin{theorem}\label{t:lim-reg2} Let the initial configuration
$X \in \Reg(\frac{K}2,\phi,K)$ and let $x'(t) < x''(t)$ be
positions of two fixed arbitrary particles at the arbitrary
moment $t$. Then the average velocity of the subconfiguration
$X_{x'(t)}^{x''(t)}$ converges to $1$ as $t\to\infty$.
\end{theorem}

\proof Denoting $\rho:=\frac{K}2$ and choosing a positive integer
$M$ we consider a configuration $^{^-M\!}X$ obtained from the
configuration $X \in \Reg(\rho,\phi,K)$ by the following
operation: for each integer $k$ we remove from the configuration
$X$ the closest from behind particle to the position $k M$. For a
given positive integer $M$ any integer $N$ may be represented as
$N=kM+l$ with $l\in\{-M,-M+1,\dots,M-1,M\}$ and $k\in\IZ^{1}$.
Then
$$ m(^{^-M\!}X_{n+1}^{n+kM+l}) =  m(X_{n+1}^{n+kM+l}) - k $$
and thus
$$ \left|\frac{m(^{^-M\!}X_{n+1}^{n+N})}N - (\rho-\frac1{M}) \right|
 = \left|\frac{m(X_{n+1}^{n+N})}N - \rho - (\frac{k}N - \frac1M) \right|
.$$
On the other hand,
$$ \left|\frac{k}N - \frac1M \right|
 = \left|\frac{k}{kM+l} - \frac1M \right|
 = \frac{l}{(kM+l)M} < \frac1N .$$
Therefore $^{^-M\!}X \in \Reg(\rho-\frac1M,~\phi+\frac1N,~K)$ and
according to Theorem~\ref{t:lim-reg1} after a finite number of
iterations $t_{c}$ the average velocity of the configuration
$T^{t_{c}}({^{^-M\!}X})$ becomes equal to $1$ (since all the
particles in this configurations are free).

Making an opposite operation, namely inserting a particle to the
configuration $X$ to the closest from behind to $kM$ empty
position for each integer $k$, we obtain another regular
configuration $^{^+M\!}X\in\Reg(\rho+\frac1M,~\phi+\frac1N,~K)$.
Again by Theorem~\ref{t:lim-reg1} after a finite number of
iterations the average velocity of this configuration becomes
equal to
$$ \frac{K}{\rho-\frac1{M}}-1 = 1 + \frac4{KM - 2} \to 1
   \quad {\rm as} \quad M \to\infty .$$

Thus both (arbitrary close as $M\to\infty$) approximations
$^{^\pm M\!}X$ to the configuration $X$ have after a finite
number of iterations (depending on $M$) the average velocity
deviating from $1$ by ${\cal O}(1/M)$. It remains to show that
the average velocity of a subconfiguration of the configuration
$X$ can be estimated from above and from below by those from
above approximations. Let $X$ and $Y$ be two configurations such
that $X(x)\le Y(x)$ for all $x$ and let $x'(t) < x''(t)$ be
positions of two fixed particles in the configuration $X$ at the
arbitrary moment $t$. Denote by $y'(t) < y''(t)$ positions of the
same particles in the configuration $Y$. Then
$$ \v(X_{x'(t)}^{x''(t)}) \ge \v(X_{x'(t)}^{x''(t)}) $$
for any moment of time $t$. Indeed, additional particles in the
configuration $Y$ present only obstacles to the motion of other
particles, thus making the average velocity slower (or at least
not faster). \qed

In the case of (space) $n$-periodic configurations numerical
examples below demonstrate a much better estimate of the
transient period: $t_c \le n-1$, but it is rather unclear if it
is possible to generalize this result for more general regular
configurations.

\begin{figure} \begin{center}
\begin{verbatim}
        t      X       v(X)           X       v(X)
        0  <0414232>   9/14       <1204440>   7/15
        1  <0142313>  10/14       <0124404>   9/15
        2  <3123131>  13/14       <4034040>  12/15
        3  <1321313>  13/14       <0430404>  12/15
        4  <3222131>  13/14       <4313040>  12/15
        5  <2222213>  13/14       <3131304>  12/15
        6  <2222222>    1         <1313133>  13/15
        7  <2222222>    1         <3131331>  13/15
        8  <2222222>    1         <1313313>  13/15
        9  <2222222>    1         <3133131>  13/15
       10  <2222222>    1         <1331313>  13/15
       11  <2222222>    1         <3313131>  13/15
       12  <2222222>    1         <3131313>  13/15
              (a)                    (b)
\end{verbatim}
\end{center} \caption{Long transient periods ($t_c=n-1$) of
$n$-periodic configurations with $K=4$ and $n=7$:
(a) $\rho=2=K/2$, \ \ (b) $\rho=15/7>K/2$. \label{ex:2}}
\end{figure}

Moreover, it turns out that even the upper (attainable) estimate
of the length of the transient period for an initial
configuration from $\Per_{\rho}(n,K)$ is not monotonous on the
length of the period $n$ and heavily depends on its parity. The
above example demonstrate the estimate $t_c \le n-1$ for odd
values of $n$. Now following mainly ideas proposed in \cite{NT}
we shall show that this estimate is rather different for even
values of the period.

\begin{lemma}\label{l:tr-per-even} Let $X\in\Per_{\rho}(2n,K)$ for
some $n\ge1$. Then the length of the transient period $t_c\le n$.
\end{lemma}

\proof We introduce an operator $G$ mapping the space of
configurations $X_{0}^{K}$ into the space of two-side sequences
of real numbers defined as follows:
$$ GX(x) := \sum_{i=0}^{x-1}X(i) - (x-1)K/2 ,$$
where we set $\sum_{i=0}^{-j}=\sum_{i=-j}^{0}$ for any positive
integer $j$.

One can easily show that for any $x\in\IZ^{1}$ we have%
$$ X(x) = GX(x+1) - GX(x) + K/2 ,$$
and if additionally $X\in\Per_{\rho}(2n,K)$, then%
\beq{e:sum-per}
    { GX(x+2n) - GX(x)
      = \sum_{i=x}^{x+2n-1}(X(i)-K/2)
      = 2nK(\rho-1/2)}%
and for any $x\in\IZ^{1}$%
\beq{e:oper-per}{G(\map X)(x) = \max\{GX(x-1),GX(x)-K/2,GX(x+1)\} .}%
This yields that
$$ G(\map^{t} X)(x)
 = \max\{\max\{GX(x-t),GX(x-t+2),\dots,GX(x+t)\}, \CR
        ~\max\{GX(x-t+1),GX(x-t+3),\dots,GX(x+t-1)\}~~ -K/2\} .$$
As usual we consider three possibilities.

(a) Assume first that $\rho<K/2$. Then from the equality
(\ref{e:sum-per}) we get that $GX(x+2n)<GX(x)$ for each $x$. Then
for $t\ge n$ we obtain
$$ G(\map^{t} X)(x)
 = \max\{\max\{GX(x-t),GX(x-t+2),\dots\},  \CR
        ~\max\{GX(x-t+1),GX(x-t+3),\dots\}~~ -K/2\} .$$
Thus $G(\map^{t+1} X)(x)=G(\map^{t} X)(x-1)$ and
$\map^{t+1}X(x)=\map^{t}(x-1)$. Substituting these equalities
into (\ref{e:oper-per}) we obtain
$$ 0 = \max\{0,~G(\map^{t} X)(x) - G(\map^{t} X)(x-1) - K/2,
               ~G(\map^{t} X)(x+1) - G(\map^{t} X)(x-1)\} \CR
     = \max\{0,~\map^{t}X(x-1)-K,~\map^{t}X(x-1)+\map^{t}X(x)-K\}, $$
from where $\map^{t}X(x)\le K-\map^{t}X(x+1)$ for any $t,x$.
Therefore for $t\ge n$ all particles in the configuration
$\map^{t}X$ are free.

(b) $\rho=K/2$. Since in this case $GX(x+2n)=GX(x)$ for all $x$
we get that for $t\ge n$
$$ G(\map^{t}X)(x)
 = \max\{\max\{GX(2),GX(4),\dots,GX(2n)\},~  \CR
         \max\{GX(1),GX(3),\dots,GX(2n-1)\} - K/2 \} $$
if $x-t$ is even and
$$ G(\map^{t}X)(x)
 = \max\{\max\{GX(1),GX(3),\dots,GX(2n-1)\},~  \CR
         \max\{GX(2),GX(4),\dots,GX(2n)\} - K/2 \} $$
otherwise. Therefore $G(\map^{t+1}X)(x)=G(\map^{t}X)(x\pm1)$ and
thus $\map^{t+1}X(x)=\map^{t}X(x\pm1)$ for all $x\in\IZ^{1}$,
which yields that $\map^{t}X\in\Free(K)$.

(c) $\rho>K/2$. This case follows from the argument applied in
the case (a), since we can consider the dual configuration
$\c{X}$, for which the density of particles is less than $K/2$.
\qed

\begin{lemma}\label{l:tr-per-odd} Let $X\in\Per_{\rho}(2n+1,K)$ for
some $n\ge1$. Then the length of the transient period $t_c\le
2n+1$.
\end{lemma}

\proof Any configuration $X\in\Per_{\rho}(2n+1,K)$ belongs also
to $\Per_{\rho}(4n+2,K)$. On the other hand the number $4n+2$ is
even and thus by Lemma~\ref{l:tr-per-even} we get the desired
estimate of the transient period. \qed

It is of interest that in the space periodic case we can give a
more detailed information about the dynamics in time.

\begin{proposition} For each configuration $X\in\Per(n,K)$ and any
integer $t\ge t_{c}=n$ the sequence $\{\map^{t}X(x)\}_{t}$ is
$n$-periodic on $t$ for each $x\in\IZ^{1}$.
\end{proposition}

\proof This result follows from the fact that for each $t$ the
configuration $\map^{t}X$ is $n$-periodic in space and by Lemmas
\ref{l:tr-per-even} and \ref{l:tr-per-odd} the length of the
transient period $t_{c}\le n$. Thus for $t\ge t_{c}$ the
configuration $\map^{t}X$ consists either of free particles (if
$\rho(X)\le K/2$) or its dual satisfies this property. Therefore
$\map^{t+n}X(x)=\map^{t}X(x)$ for any $x\in\IZ^{1}$. Notice that
this period in time might be not minimal (which can be as small
as 2). \qed

Observe that this construction heavily depends on the periodic
space structure of the configurations, which rules out the
generalization for a more general situation.

\section{On the chaoticity of the dynamics}\label{s:chaos}

In the previous sections it was shown that for sufficiently large
time the dynamics occurs either in $\Free(K)$ or in
$\c{(\Free(K))}$, i.e. the corresponding dual configurations
belong to this space. Therefore to study asymptotic (as time goes
to $\infty$) properties of the dynamics we consider its
restriction to the space of configurations of free particles
(which contains the union of all attractors of the map $T$
restricted to the set of regular configurations with the density
less than $K/2$). The following result shows that this map is
chaotic in the sense that its topological entropy (see
definitions in \cite{ME}) is positive.

\begin{theorem}\label{t:chaos}
$h_{{\rm top}}(T,\Free(K))
 = \ln\left(\frac{2(K+1)}\pi + \frac1\pi
 + \frac{R(K)}{(K+1)^{2}} \right) > 0$ for any $K\in\IZ_{+}^{1}$.
The remainder term $R(K)$ above satisfies the inequality
$|R(K)|\le2$.
\end{theorem}

\proof All the particles in configurations on the largest (i.e.
containing all others) attractor are free, i.e. $X(x)+X(x+1)\le
K$. Thus the action of the map is equivalent to the right shift
map with the upper triangular transition matrix (i.e. all
elements in the first line are 1, all but the last are 1's in the
2nd line, etc.). It is well known (see, for example, \cite{ME})
that the logarithm of the largest eigenvalue of this matrix gives
the topological entropy of the right shift map and thus the
topological entropy $h_{{\rm top}}(T)$ of the traffic flow.
Therefore the representation of the largest eigenvalue of the
transition matrix which we shall give below finishes the proof.
\qed

To simplify the notation we denote $N:=K+1$. Let $A(N)=(a_{ij})$
be the $N\times N$ left triangular matrix, i.e. $a_{ij}=1$ for
all $i+j\le N+1$ and $a_{ij}=0$ otherwise. This is a nonnegative
symmetric matrix, therefore its spectrum belongs to the real line
and its largest eigenvalue $\la_{\max}(A(N))$ is positive.

\begin{theorem}\label{t:detailed-eig} $\la_{\max}(A(N))=\frac2\pi N +
\frac1\pi + \frac{R}{N^{2}}$, where the remainder term
$|R|=|R(N)|\le2$.
\end{theorem}

\proof For an integrable function $f\in\IL^{2}$ consider the
operator
$$ Lf(x) := \int_{0}^{1-x}f(s)~ds .$$
According to \cite{Bl-sp} eigenvalues of the operator $L$ ordered
by their moduli are equal to
$$ \la_{k}:=\frac{(-1)^{k+1}}{(k-1/2)\pi} ,$$
while $\Ee(x):=\cos(\frac\pi2 x)$ is the eigenfunction
corresponding to the leading eigenvalue $\la_{1}$. For simplicity
we shall use the notation $A:=A(N)$, $\la:=\la_{1}=\frac2\Pi$,
$\Ee_{k}:=\Ee(k/N)$, and $\ep=\frac1{N}$. Now since the function
$\Ee(x)$ is analytical, decreases monotonically, and its second
derivative satisfies the inequality
$|\frac{d^{2}\Ee(x)}{dx^{2}}|<\frac{\pi^{2}}4$, it follows
that for each $k=1,2,\dots,N$ we have%
\beq{e:matr-oper-1}{
 (A\Ee)_{k} = \int_{0}^{1-k/N}\Ee(s)~ds
          + \frac1{2N}\sum_{i=0}^{N-k}
                      \left(\Ee(\frac{i}N) - \Ee(\frac{i+1}N)\right)
          + \frac{R_{1}}{N^{2}} ,}%
where the remainder term
$|R_{1}|=|R_{1}(N)|\le\frac{\pi^{2}}{4\cdot6}<1$. Thus,
introducing the operator $Gf(x) := f(0) - f(1-x)$, we rewrite the
last equality as
$$ (A\Ee)_{k} = L\Ee(k/N) + \frac\ep2 G\Ee(k/N) + R_{1}\ep^{2} .$$
On this step our aim is to show that there exists a function
$g\in\bC^{1}[0,1]$ orthogonal to $\Ee$ (which means that $\int
g\cdot\Ee=0$) such that the following relation holds true:%
\beq{e:matr-oper-2}{
 (L + \frac\ep2 G)(\Ee + \frac\ep2 g)
 = (\frac2\pi + \ep\frac1\pi)(\Ee + \frac\ep2 g) + R_{3}\ep^{2} ,}%
where again $|R_{3}|\le1$.

Since the operator $L$ is symmetric the orthogonal complement to
the function $\Ee$ is invariant with respect to $L$. Thus for
some constant $\alpha$ and a bounded function $h\in\bC^{1}$
independent on $\ep$ and orthogonal to $\Ee$ we have
$$ G\Ee = \alpha \Ee + h .$$
Let us calculate the constant $\alpha$. Since $h$ is orthogonal
to $\Ee$, then multiplying the both hands of the previous
equality by $\Ee$ and integrating (observe that
$\int_{0}^{1}\Ee\cdot h=0$) we get
$$ \int_{0}^{1} \Ee(x)\cdot(\Ee(0) - \Ee(1-x))~dx
 = \alpha \int_{0}^{1}\Ee^{2}(x)~dx .$$
On the other hand,
$$ \int_{0}^{1}\cos^{2}(\frac\pi2 x)~dx = \frac12, \qquad
   \int_{0}^{1}\cos(\frac\pi2 x)~dx = \frac2\pi, \CR
   \int_{0}^{1}\cos(\frac\pi2 x)~\cos(\frac\pi2 (1-x))~dx
 = \frac12 \int_{0}^{1}\sin(\frac\pi x) = \frac1\pi . $$
Thus
$$ (\frac2\pi - \frac1\pi) = \frac12 \alpha,$$
from where $\alpha=\frac2\pi$.

Therefore for any function $g\in\bC^{0}$ we have
$$ (L + \frac\ep2 G)(\Ee + \frac\ep2 g)
 = L\Ee + \frac\ep2 Lg + \frac\ep2 G\Ee + \frac{\ep^{2}}4 Gg
 = \left(\frac2\pi + \frac\ep2 \frac2\pi\right)\cdot \Ee
 + \frac\ep2(h + Lg) + \frac{\ep^{2}}4 Gg .$$
Comparing this relation with the equality (\ref{e:matr-oper-2})
we come to the conclusion that $h + Lg = \frac2\pi g$ or
$g:=(L-\frac2\pi)h$. Notice that the right hand side of the last
expression makes sense since $h$ is orthogonal to $\Ee$. Thus
$$ g(x) = (L - \frac2\pi)^{-1}\left(\Ee(0) - \Ee(1-x)
             - \frac1\pi \Ee(x) \right) .$$
From the first two leading eigenvalues, we deduce that the norm
of the operator $L - \frac2\pi$ restricted to the orthogonal
complement to the leading eigenfunction can be estimated from
above by $\frac2\pi - \frac2{3\pi}=\frac4{3\pi}$. Therefore
$$ |g| \le \frac4{3\pi}(1 - \frac1\pi), \qquad
   \frac14|Gg| \le \frac{3\pi(1+\frac1\pi)}{4\cdot4} < 1 .$$

Combining above estimates we deduce that there exist two vectors
$v,\xi\in\IR^{N}$ such that%
$$ Av = (\frac2\pi N + \frac1\pi)v + \xi, \qquad
  |\xi|\le\frac2{N^{2}}|v| ,$$
which yields the statement of Theorem by the following
a'posteriori matrix perturbation argument \cite{Pa}. Let the
equality $Av = \mu v + \xi$ be satisfied for a symmetric matrix
$A$, two vectors $v,\xi\in\IR^n$ with $||\xi|| \le \ep||v||$ and
a scalar $\mu$. Then the closest to $\mu$ eigenvalue $\la$ of the
matrix $A$ satisfies the inequality $|\la-\mu|\le\ep$.

Indeed, for $\mu=\la$ the inequality becomes trivial, while
otherwise
$$ ||v|| \le ||(A - \mu I)^{-1}|| \cdot ||(A - \mu I)v||
 = \frac1{|\mu-\la|}\cdot ||\xi|| .$$
Thus $|\mu-\la| \le \frac{||\xi||}{||v||} \le \ep $. \qed

As we already mentioned this result corresponds to the steady
states of our model for the case of initial regular
configurations with `low' traffic. Notice that in the opposite
case of `high' traffic $\rho>K/2$ each jammed configuration is in
one-to-one correspondence with its dual one, for which
$\rho<K/2$. Therefore the statement similar to
Theorem~\ref{t:chaos} holds in this case as well, moreover
$$ h_{{\rm top}}(\map,\Free(K)\cup\c{(\Free(K))})
 = h_{{\rm top}}(\map,\Free(K)) .$$
Observe that for any positive integer $K$ the topological entropy
for the map $T$ is strictly positive, which yields the chaoticity
of the map.

\section{Statistics of typical configurations}\label{s:statistics}

In this section we shall derive statistical information about
typical configurations of particles. For configurations
$X\in\Free(K)$ let us denote by $S(n,K)$ the total number of
different subconfigurations $X_{1}^{n}$ of length
$n\in\IZ^{1}_{+}$.

\begin{lemma}\label{l:stat-1} $S(n,K) = \la_{\max}^{n}(A(K+1))
+ o(\la_{\max}^{n}(A(K+1)))$, where $A(N)$ is $N\times N$ left
triangular matrix.
\end{lemma}

\proof Denote by $S_i(n,K)$ the number of subconfigurations of
length $n$ consisting of only free particles and starting with
the symbol $i\in\{0,1,\dots,K\}$. Then we have the following
recurrence relation:
$$ S_i(n+1,K) = \sum_{j=0}^{K-i} S_j(n,K) .$$
The number of particles in each site of a configuration $X$ may
vary from $0$ to $K$, i.e. it may admit $N:=K+1$ different
values, and the only additional relation that should be satisfied
is
$$ X(x) + X(x+1) \le K \quad \forall x\in\{1,2,\dots,n\} .$$
Therefore these configurations are completely described by
$N\times N$ left triangular transition matrix $A=A(N)$. Thus we
get $S(n,K) = \sum_{i=0}^K S_i(n,K)$, from where and from
Theorem~\ref{t:detailed-eig} the statement of Lemma follows. \qed

We shall say that a subconfiguration is {\em blocking
(non-blocking)} if it contains (not contains) the symbol $K$.
Then the number of non-blocking subconfigurations of length $n$
is equal to $S(n,K-1)$. The fraction of blocking
subconfigurations of length $n$ is equal to
$$ \frac{S(n,N) - S(n,N-1)}{S(n,N)} = 1 - \frac{S(n,N-1)}{S(n,N)} .$$
Applying the asymptotic representation for the leading eigenvalue
of the matrix $A(N)$ we get the following estimate:
$$ S(n,N) = \left(\frac2\pi N + \frac1\pi + \om(\frac1N)\right)^{n} .$$
Therefore
$$ \frac{S(n,N-1)}{S(n,N)}
 = \left(\frac{\frac2\pi (N-1) + \frac1\pi + \om(\frac1N)}
              {\frac2\pi N + \frac1\pi + \om(\frac1N)}\right)^{n}
 = \left(1 - \frac1N + \om(\frac1N)\right)^{n} .$$

To derive a more deep information about the statistics and to be
able to deal with periodic configurations one can use an approach
based on Markov chain approximations.

For for a given integer $i\in\{0,1,\dots,K\}$ and a configuration
$X\in\Free(K)$ denote by $\bar\pi_{i}(X_{1}^{n})$ the fraction of
sites $x\in\{1,2,\dots,n\}$ where $X(x)=i$, i.e.
$$ \bar\pi_{i}(X_{1}^{n}) := \frac1n~ \#\{x\in\{1,\dots,n\}:~~X(x)=i\} ,$$
while by $\bar\pi_{i}^{(n)}$ we denote the average of these
fractions over all possible different subconfigurations of free
particles of length $n$:
$$ \bar\pi_{i}^{(n)}
 := \frac{\sum_{X\in\Free(X)}\bar\pi_{i}(X_{1}^{n})}{S(n,K)} .$$

\begin{lemma}\label{l:mark-appr}
$\bar\pi_{i}^{(n)} \to \frac{K+1-i}{K+1}\frac2{K+2}$ as
$n\to\infty$.
\end{lemma}

\proof Continuing the same argument as in the proof of
Lemma~\ref{l:stat-1} we see that each configuration can be
considered as a realization of a Markov chain with $N=K+1$ states
numbered as $0,2,\dots,K$ and the following transition
probabilities:
$$ p_{ij} := \function{{\cal P}\{X(x+1)=j~{\bf |}~X(x)=i\} = \frac1{K-i+1}
                  &\mbox{if } j\le K-i \\  0 &\mbox{otherwise}.} $$
Clearly the $N$-th power $P^{N}$ of the transition matrix
$P=(p_{ij})$ is strictly positive and thus the Markov chain is
ergodic. Denote by $\pi_i$, $i\in\{0,\dots,K\}$ its stationary
probabilities, i.e. the probability to have $X(x)=i$. Then these
quantities should satisfy the following system of equalities:
$$ \pi_i = \sum_{j=0}^{K-i} \frac{\pi_j}{K-j+1} ,$$
from where
$$ \pi_i - \pi_{i+1} = \frac{\pi_{(K-i)}}{i+1} $$
for all $i=0,1,\dots,K$. Solving the last system of difference
equations we get
$$ \pi_i = \frac{K+1-i}{K+1} \pi_0 ,$$
and eventually (since they sum up to 1) we come to
$$ \pi_i = \frac{K+1-i}{K+1}\frac2{K+2} .$$
\qed

To study statistics of $n$-periodic configurations of free
particles one should take into account that there is an
additional constraint: $X(1)+X(n)\le N$. Denote by $\tilde
S(n,N)$ the total number of subconfigurations of length $n$
consisting of free particles and satisfying this constraint.
Therefore the fraction of nonadmissible $n$-periodic
configurations (which do not satisfy above constraint)
$$ \sum_{i=0}^{K}\pi_{i} \sum_{j=K-i}^{K}\pi_{j}
 = \sum_{i=0}^{K}\pi_{i} \left(1 - \sum_{j=0}^{K-i}\pi_{j} \right) $$
is asymptotically (for large $K$) equal to
$$ \int_{0}^{1}\pi(x)~\int_{1-x}^{1}\pi(y)~dy~dx = \frac16 ,$$
where $\pi(x):=2(1-x)$ corresponds to the limit (as $K\to\infty$)
distribution.

Using these statistics one can easily obtain all correlation
functions and to study large deviations. For example, we get that
the fraction of blocking configurations in $\Free(K)$ is equal to
$\pi_{K}=2/((K+1)(K+2))$, which in terms of traffic estimates how
often the road with $K$ lanes is completely blocked by moving
cars.

%%%--------------------------------------------------------%%%%
\newpage%
{\small%

}

\end{document}